\documentclass[]{spiejour}  
\usepackage[]{graphicx}
\usepackage{epstopdf,mathrsfs,amsmath,amsfonts,textcomp,amssymb}

\newtheorem{postulate}[theorem]{Postulate}

\title{An electromagnetic model for biological tissue}

\author{Luisiana X. Cundin}

\authorinfo{Luisiana X. Cundin is a contractor for Naval Medical Research Unit - San Antonio (NAMRU-SA), contracted through Conceptual Mindworks, Inc.}

\begin{document} 
  \maketitle 

\begin{abstract}
This essay is a recapitulation of an earlier Kramers-Kr\"{o}nig analysis of biological tissue, published in 2010 \cite{2010arXiv1010.3752C}. The intent is to both complement and bolster the antecedent analysis by furnishing supplemental clarification on the electromagnetic model employed and provide further technical details. Biological tissue is modeled a dielectric embedded in a conductor, which necessitates using both a quotient and product subspace to form a complete topological cover space. Discerning a suitable quotient enables decisive separation of tissue's dielectric behavior from excess conductivity. The residual dielectric behavior revealed conforms with expectations based upon electromagnetic theory and those commonly held for dielectric materials. An appreciation for conflicting experimental absorption measurements spanning the optical spectrum, reported for biological skin, engenders conceptualizing a damage dependent attenuation coefficient. Lastly, descriptive codes are provided for numerical algorithms implemented in the original analysis \cite{2010arXiv1010.3752C}. 
\end{abstract}

\section{Introduction}
Over one hundred years of empirical research has supported high permittivity measurements for biological tissue, and this would \emph{prima facie} not be a source of contention, if it were not for biological tissue being classified a dielectric. Dielectric materials are polarizable insulators and, by definition, insulators are poor conductors; yet, experimentation has proven biological tissue a fairly good conductor, of course, tissue is still a poor conductor relative to a metal. Pure liquid water is a classic dielectric material and known for its large dielectric constant; yet, experiment supports a dielectric constant for biological tissue some five orders of magnitude greater than that of pure liquid water. Augmented polarizability for tissue is associated with electric double layers surrounding cellular membranes. Electrolytic solutions, such as saline solutions, admit a depressed polarizability relative to pure liquid water and is due to shielding by ions; albeit, the same solvated ions increase the conductivity of the media, attributable to electric double layers formed by solvated ionic cages. 

What is meant by 'tissue' is in truth many different types of tissue found within a biological body. The specific permittivity measured for each respective tissue type can vary, due to variations in chemical composition, density, salinity, \&c. The intrinsic properties of tissue types can vary across the taxonomic hierarchy for biology, including lower subclasses. In addition, the resultant dielectric behavior can be further affected by such variables as age, disease, environment and so on. With so many dependent variables defining 'tissue', its intrinsic properties, its \emph{resultant} dielectric behavior; it seems any unified treatment of tissue would be prohibited. 

A biological body has a fractal nature, starting with the cellular structure at the mesoscopic scale, the cellular structure subdivides space into repeated isolated enclaves translated throughout the entire biological body; further still, the body is broken into several distinct organs and this lends to many particular tissue types. There are several different cell types, each based upon the intended function for that cell; furthermore, this is throughout the whole tissue field. Organs are a collection of cells with the same or similar functionalities, e.g. liver. A biological cell differentiates space and is delineated by the plasma membrane, which separates the entire space into two disjoint subspaces, namely, the intra and extra-cellular subspaces.

The primary implication, of present interest concerning the plasma membrane, despite having multiple functionalities, is the creation and maintenance of an electric potential across the membrane. Biological activity is responsible for the maintenance of the membrane potential, which is typically maintained somewhere between -50 to -170 millivolts; specific potentials depend upon the function of the cell and many other factors. Maintenance requires the continual ejection of positive ions from the intra-cellular region, for smaller ions, like sodium ions, may slip through the membrane with relative ease. The electric potential across the membrane is negative, indicating the direction of the potential is pointed inward towards the center of the cell. Positive ions, such as sodium ions, congregate near the outer surface of the plasma membrane. Negative ions, such as chloride ions, congregate around the cell, somewhere in the vicinity to balance charge. The separation of charged ions around the outer portion of the cellular membrane creates a thin conductive layer referred to as either the 'Stern layer', 'electric double layer' or, as simply, the 'double layer' \cite{Interfacial}. 

An electric double layer is an interfacial phenomenon occurring in fluids and consists of like charged particles collecting on the surface of some host particle; while, opposingly charged particles congregate in the vicinity to maintain a balance of electric charge. There is a net charge that is null for the entire solution, but locally, there are sustained electric potentials, as in the case of cellular membranes. The layer of oppositely charged particles is referred to as the diffuse layer; this layer is comprised of charged species freely moving in and around the vicinity of the host particle, yet the net effect is to maintain a proper charge balance. This phenomenon occurs in many different fluids, such as air, liquids and even molten metals. 

If it were not for the cellular structure, susceptible augmentation of tissue would not be expected and experiment bears this out. The dielectric constant for tissue fluid, that is fluid absent any cellular structure, generally shows a maximum dielectric constant in the neighborhood of two orders of magnitude, roughly that of pure liquid water; conversely, tissue containing cells, such as liver or gallbladder, exhibit extremely large dielectric constants, typically somewhere in the neighborhood of six to eight orders of magnitude. The increase in polarizability of biological tissue is directly attributable to the double layer, where charged particles are free to migrate to antipodal points on the surface of a cell. Given the typical diameter of a cell, \emph{circa} ten microns, it is not surprising to see such large dielectric constants. 

Besides contributing to the creation of an electric double layer surrounding the cellular membrane, solvated ions also disassociate in solution and thereby give up a weakly held valence electron, called a solvated electron. There are many descriptions for the electric properties of electrolytic solutions, but the electric double layer model can be used with great effect. The ion's potential preferentially orients water molecules to surround the ion in a quasi-lattice structure, referred to as a solvated cage; the solvated cage structure itself constitutes an electric double layer. The advent of solvated cages throughout the biological fluid provides a quasi-periodic structure allowing the propagation of valence electrons throughout the media. As a consequence, electrolytic solutions admit higher conductivity than pure liquid water, e.g. saline water and biological tissue. The induced conductive property is proportional to the ionic concentration and it is known that at saturation the conductive properties saturate; thus, at some point, there is no increase in conduction, because there is no longer an increase in solvated electrons, i.e. there is no further dissolution of salt material.  

Electric double layers account for excess conductivity in biological tissue, but how does excess conductivity relate to embedded dielectric materials? Given that dielectrics are poor conductors, should bulk conductivity values measured for tissue be ascribed to all embedded dielectric materials? If dielectric materials are poor conductors, then shouldn't excess conductivity exhibited by tissue be considered a separate phenomenon all together? Considering all embedded dielectric materials as just that, embedded materials within a fluid conductor, leads to the postulate that tissue's dielectric property is a topological quotient subspace. In this model, dielectric materials are considered to be a closed set, that is, some particle with finite dimensions and physical extent; hence, the outer boundary of such particles come into contact with the surrounding media, which in the case of tissue is a conductive media. Based upon electromagnetic boundary conditions along an interface between two separate mediums, an equivalence class is defined to maintain continuity of all fields across interfaces, known collectively as \emph{field continuity conditions}. In the case of tissue, the \emph{field continuity conditions} impose a quotient subspace topology with regard to excess conductivity and a product subspace with regard to all embedded dielectric materials. Once a suitable quotient is realized, the excess conductivity associated with electric double layers can be parsed out, rendering the covered dielectric behavior exposed. The residual dielectric behavior aligns with common expectations for dielectric materials; moreover, predicted influences from such processes as electrophoresis and ionic shielding become clearly evident. 

The combined or \emph{effective} intrinsic property of a heterogeneous material is often modeled using Maxwell's 'mixing rules', where such rules are subordinate to a wider theory called the \emph{effective medium approximation}. The theory states the combined or overall intrinsic property of a heterogeneous material is equal to a weighted sum of all the constituents making up that compound. This is akin to a \emph{resultant} property; wherefore, the resultant vector in a vector field is dominated by the largest magnitude and direction of all the vectors in that field. Considering the chemical makeup of tissue, made primarily of liquid water, it is certainly reasonable to expect the dielectric behavior of tissue to mimic that of pure liquid water. Admittedly, there should be deviations from pure liquid water; but, it should equally be possible to account for all such deviations through common understanding of electromagnetic phenomena. 

After applying a suitable quotient transforming experimental conductivity values for biological tissue, expected absorption characteristics for tissue are clearly visible in the residual absorption curve. Such characteristics as the expected contraction of the thermal resonance peak, the raised absorption peak due to electrophoresis and the overall resemblance to pure liquid water are all displayed nicely enough. The expected absorption characteristics of tissue has been discussed in the original analysis, but the transformation of experimental radio frequency measurements requires further clarification. 

Expectations are based upon electrolytes in solution, which should increase the absorption for slowly varying fields, increasing monotonically to peak around the resonant frequency associated with electrophoresis. Another consequence of electrolytes in solution is the shielding of water molecules from external fields; thereby, forming 'bound' water. Bound water refers to the phenomenon whereby water molecules are impeded in their free rotation and leads to contraction of the thermal resonance peak, that is, dilation of the relaxation time. A reduction in absorption is expected for those frequencies far removed from the resonant frequency; yet, in the neighborhood of the resonance frequency, rotational excitement of water molecules is expected to occur in spite of ionic shielding. 

The discussion above covers mostly the lower frequency range, in the case of the optical frequency range, specifically, the near-infrared, visible and near-ultraviolet spectrum, an overall increase in absorption of tissue is justly expected, for common experience proves tissue is not transparent at these frequencies, unlike pure liquid water. An increase in absorption is not what is at contention, contention exists over exactly how much of an increase is expected; worse yet, experimentation has shown two separate magnitudes, generally, for tissue's electromagnetic attenuation. Even though there are many differences in experimental techniques used to measure the optical property of tissue, the one characteristic common to each separate group of measurements is that the group admitting a lower absorption is, generally speaking,  derived from \emph{in vivo} samples, while the other group is derived from \emph{ex vivo} sample preparation. As a consequence, it is postulated that the differences in absorption measurements are attributable to tissue denaturation and coagulation, that is, the difference is between living and dead tissue. 
\newpage

\section{Theoretical models}\label{theory}
In conjunction with Maxwell's electromagnetic field equations, appropriate boundary conditions must be imposed at all interfaces to realize a complete solution. Each constitutive relation relates a fundamental field to its corresponding perturbation field, e.g. the displacement field $\vec{\mbox{D}}$ is related to the electric field $\vec{\mbox{E}}$ through the permittivity $\hat{\epsilon}$ thus $\vec{\mbox{D}}=\hat{\epsilon}\,\vec{\mbox{E}}$. At an interface possessing charge density, the \emph{field continuity conditions} impose certain constraints on the fundamental, as well as, perturbation fields. At all interfaces, each adjoining region in a domain must be glued along the boundaries, which results in a quotient mapping for each interface, sending each boundary point to its equivalence class \cite{Lee}. An example of the boundary principle is a finite dielectric body surrounded by a vacuum, to attain the effective dielectric constant, one must divide the permittivity of the dielectric material by the vacuum permittivity thus $\hat{\epsilon}_{\mbox{\scriptsize{\it{eff}}}}=\hat{\epsilon}/\epsilon_{\mbox{\scriptsize{0}}}$ \cite{Landau,Jackson}. 

Biological tissue experiences a conductivity augmented from pure liquid water, attributable to both electrolytes in solution and the double layer surrounding each biological cell; as a consequence, it is postulated that all dielectric materials in tissue are embedded in a conductive media. This means we must adjoin to the outer boundary of all dielectric regions the inner boundary of a circumscribing conductor. 
\begin{postulate}[dielectric embedded in a conductor]\label{quotientconductor}
Because of excess conductivity exhibited by biological tissue, all embedded dielectric materials must be surrounded by a conductive media; as a consequence, each constitutive relation for an embedded dielectric ($\mbox{\scriptsize{d}}$) is divided by that of a conductor ($\mbox{\scriptsize{c}}$).
\begin{equation}
 \sigma^\prime_{\mbox{\scriptsize{eff}}}=\sigma^\prime_{\mbox{\scriptsize{d}}}/\sigma^\prime_{\mbox{\scriptsize{c}}}
\end{equation}
Where the ratio equals the effective ($\mbox{\scriptsize{eff}}$) constitutive relation, in this case, conductivity. 
\end{postulate}

The process of equating fields along an interface is a quotient mapping that sends each boundary point to its equivalence class. For example, the quotient mapping, $\hat{\epsilon}:\vec{\mbox{D}}\mapsto\vec{\mbox{D}}/\hspace{-3pt}\sim$, represents a partition of $\vec{\mbox{D}}$ and together with the quotient topology determined by the constitutive relation is called the \emph{quotient} space of the perturbation field. This process is repeated for all other fields, such as the induction $\vec{\mbox{B}}$, magnetic $\vec{\mbox{M}}$ and electric $\vec{\mbox{E}}$ fields; moreover, the field's orientation, either tangential or normal to the surface, must also be considered.

In the case of dielectric materials comprising tissue, they do not contain one another; as a result, the \emph{effective} constitutive relation would form a product space. Each material would form a basis $\mathscr{B}$ for the resulting topology $\mathscr{X}$, where the union of all constitutive relations form an effective mapping \cite{Lee}. In keeping with the \emph{effective medium approximation}, a weighted sum is applied to each mapping in proportion to each material's concentration.
\begin{postulate}[product space for dielectrics]\label{productdielectric}
A finite set of dielectrics form a basis $\mathscr{B}$ for the topology $\mathscr{X}$ known as biological tissue. The union of all such mappings (constitutive relations) form a product topology for tissue, where each mapping ($\hat{\epsilon}_i$) is weighted ($a_i$) proportionally to each constituent's aliquot part.
\begin{equation}
 \bigcup_{i}a_i\hat{\epsilon}_i\subseteq \hat{\epsilon}_{\mbox{\scriptsize{\textit{d}}}}
\end{equation}
Where each mapping $\hat{\epsilon}_i:\vec{\emph{\mbox{E}}}_i\mapsto \vec{\emph{\mbox{D}}}_i$ forms a subset basis for the topology $\mathscr{X}$, the union of all weights ($a_i$) is unity and the representative constitutive relation chosen is permittivity.  
\end{postulate}

Modification of the basis set $\mathscr{B}$ amounts to a shift in the partial molar concentration of participating constituents; the resulting topology would change proportionately. All tissue types are spanned by suitable variation of all basis sets comprising biological tissue. Additionally, variation in the intersection of all possible basis sets, that is, introducing or removing particular chemical constituents, would equally change the resultant topology. Postulate \ref{productdielectric} is an equivalent statement of the \emph{effective medium approximation} theory.

With respect to conductivity, it is a topological property stemming not from a product space, but a quotient space; thus, resolving the dielectric behavior from excess conductivity attributable to electrolytes and electric double layers would require multiplication by a quotient. Once the quotient is accomplished, the revealed dielectric behavior for biological tissue could theoretically be deconstructed by aliquot subtraction of each chemical constituent comprising the mixture. It is at this point of the analysis that one may apply the concept of an \emph{effective medium approximation} and predict the dielectric constant for nearly static fields, adequately explain observed deviations of the absorptive behavior from pure liquid water and, finally, provide some insight in to how electromagnetic fields might interact with biological tissue. 

In addition to contraction of the thermal resonance peak, another absorption peak is expected to be raised due to an increase in absorption from electrophoresis, which is the transportation of ionic species through a liquid. Consider a static field, we can imagine the nucleus moving through the medium under the force of an applied field, similarly, the electron cloud would move opposingly to the direction of the nucleus. This electronic displacement sets up a local electric potential, where the attraction between the electron cloud and nucleus retards the movement of the ion through the medium. Because of these forces, ion mobility is limited in speed and therefore represents a loss of energy to the medium. Consider now an oscillating field, we can imagine now the nucleus and electron cloud shifting relative positions proportionate to the applied field; furthermore, at some frequency, both will oscillate so rapidly as to vibrate the nucleus around some central point in space. Assume the displacement of the nucleus is on the order of a nucleic width, then the applied field creates a resonant frequency associated with electrophoresis phenomenon.

Given a set of ionic species, there will be a distribution of corresponding ionic mobilities, with an average velocity of $10^{-8}$ meters per second \cite{Atkins}. The resonant frequency $f$ can be calculated by assuming the oscillations to occur in one-dimension, with velocity $v$ and width $d$; then, there will be $N$ normal modes, where $N$ equal to unity corresponds to the fundamental frequency, see equation (\ref{resonance}). Given that the typical width of an ionic nucleus is on the order of femtometers, a fundamental frequency can be calculated as roughly 10 megahertz.
\begin{equation}\label{resonance}
 f=\frac{Nv}{2d},N=\{1,2,3,\ldots\}
\end{equation}

For an externally applied field, as the frequency of oscillation increases, absorptive loss associated with electrophoresis is expected to monotonically rise and peak around 10 megahertz. Given that a distribution of ionic species are present in biological tissue, with corresponding distributions in ionic mobilities, velocities and nucleic widths, the absorption curve associated with electrophoresis should be broad and smooth, for there are in addition, a range of normal modes to be considered as well. 

Lastly, damaged tissue is characterized by unfurled proteins and this is expected to increase the opacity of tissue. Experiment has shown in the near-infrared, visible and near-ultraviolet spectrum that tissue exhibits either very low absorption of electromagnetic energy, if the sample tested is \emph{in vivo}; in contrast, if \emph{ex vivo} sample preparation is used, much higher absorption is generally measured. It is postulated that the range measured for tissue's electromagnetic attenuation is caused by damage related to temperature, for temperature directly affects the degree of coagulation, denaturation or unfolding of proteins within tissue, thus increasing the opacity of biological tissue.
\begin{postulate}[damage dependent absorption coefficient]\label{coefficient_theory}
There is a mapping $\mu_a$ that maps tissue attenuation and is a function of temperature ($T$), laser power density ($P_d$) and possibly other parameters, including time ($t$). The mapping sends the absorption ($k_1$) for undamaged tissue continuously to the absorption ($k_2$) of damaged tissue. 
\begin{equation}
 \mu_a\hspace{-2.5pt}\left(\mbox{T},\mbox{P}_{\mbox{\scriptsize{d}}},\ldots;\mbox{t}\right):k_1\mapsto k_2
\end{equation} 
\end{postulate}
\newpage

\section{Technical notes}
The intent of the present essay is not to relive all the details relayed in the original analysis, but to complement and bolster the antecedent analysis \cite{2010arXiv1010.3752C}. Three main topics are to be covered in what follows: the implications of Postulate \ref{quotientconductor} are explored for radio frequency conductivity data, then a brief discussion of optical frequency data shows that the penetration depth is a function of time, finally, four numerical algorithms are briefly discussed. For the sake of brevity, formulas already extant in the original paper will be referenced throughout what follows.

One definitive source suffices for experimental conductivity data covering the radio frequency range and it is to this set of conductivity data a suitable quotient must be multiplied in order to parse out the covered dielectric behavior of tissue \cite{Gabriel:AFRL}. All relevant constitutive relations are linked together into one succinct formulae, the complex index of refraction $\hat{\mbox{N}}$, see equation (2) found in the original analysis \cite{2010arXiv1010.3752C}.

Using the definition of $\hat{\mbox{N}}$, equation (\ref{approximation}) can be deduced with the aid of Minkowski's inequality, Theorem 2.1 in the original paper, and approximates the absorption behavior in the lower frequency range \cite{2010arXiv1010.3752C}. Equation (\ref{approximation}) states the absorption $k$ of electromagnetic energy is approximately equal to the square root of the relative conductivity $\sigma_r^\prime$, divided by the angular frequency $\omega$, multiplied by the algebraic term ($4\pi+1$). It is by this formulae absorption is often related to conductivity in the lower frequency range.
\begin{equation}\label{approximation}
 k\approx\sqrt{\frac{\sigma_r^\prime}{\omega}(4\pi+1)}
\end{equation}

Now, the entire biological body is a finite body and is eventually surrounded by air; therefore, the conductivity for the conductor $\sigma_{\mbox{\scriptsize{\textit{c}}}}^\prime$, found in Postulate \ref{quotientconductor}, is relative to a vacuum. For the conductor contains all dielectrics in tissue, but air contains the conductor. Starting with the relative conductivity term in equation (\ref{approximation}), it is possible to trace this term through and finally relate it directly to the experimental conductivity $\sigma^\prime_{\mbox{\scriptsize{exp}}}$ measured for tissue, \emph{viz}. :
\begin{equation}\label{equality}
\frac{\sigma^\prime_r}{\omega}\equiv\frac{1}{\epsilon_{\mbox{\scriptsize{0}}}\omega}\sigma^\prime\equiv\frac{1}{\epsilon_{\mbox{\scriptsize{0}}}\omega}\sigma^\prime_{\mbox{\scriptsize{\textit{eff}}}}\equiv\frac{1}{\epsilon_{\mbox{\scriptsize{0}}}\omega}\frac{\sigma_{\mbox{\scriptsize{\textit{d}}}}^\prime}{\sigma_{\mbox{\scriptsize{\textit{c}}}}^\prime/(\epsilon_{\mbox{\scriptsize{0}}}\omega)}\equiv\frac{\sigma_{\mbox{\scriptsize{\textit{d}}}}^\prime}{\sigma_{\mbox{\scriptsize{\textit{c}}}}^\prime}\equiv\sigma^\prime_{\mbox{\scriptsize{exp}}}
\end{equation}

Thus, the experimental conductivity $\sigma^\prime_{\mbox{\scriptsize{exp}}}$ values reported for biological tissue is, by Postulate \ref{quotientconductor}, equated as the ratio of the conductivity for a dielectric over that of a conductor. Since the dielectric behavior of biological tissue is what is of interest, a quotient is required to remove the conductivity ($\sigma_{\mbox{\scriptsize{\textit{c}}}}^\prime$) term found in the denominator; furthermore, this term represents the conductivity of the inner edge of a conductor that is intended to be glued to the outer edge of a dielectric material. 

The conductivity attributed to any metal is truly confined to a thin region skirting the outer edge of the metal and is known as the skin depth. As for inside a metal, the minimum conductivity possible is equal to the \emph{vacuum displacement current}; because, even in a vacuum, a current exists, also, the inner conductivity of a metal would rise from the floor value proportional to how poor a conductor the metal should be.

If the minimum conductivity is assumed along the inner edge for all electric double layers, then the \emph{vacuum displacement current} becomes the suitable quotient that must be applied to experimental conductivity measurements. Thus, multiplying the experimental conductivity $\sigma^\prime_{\mbox{\scriptsize{exp}}}$ values reported for biological tissue by the \emph{vacuum displacement current}, then substitution of this product into equation (\ref{approximation}) yields the following:
\begin{equation}\label{approx2}
 k\approx\sqrt{\epsilon_{\mbox{\scriptsize{0}}}\omega\sigma^\prime_{\mbox{\scriptsize{exp}}}(4\pi+1)}
\end{equation}

\begin{figure}[t]
\centering
 \includegraphics[width=4.0 in]{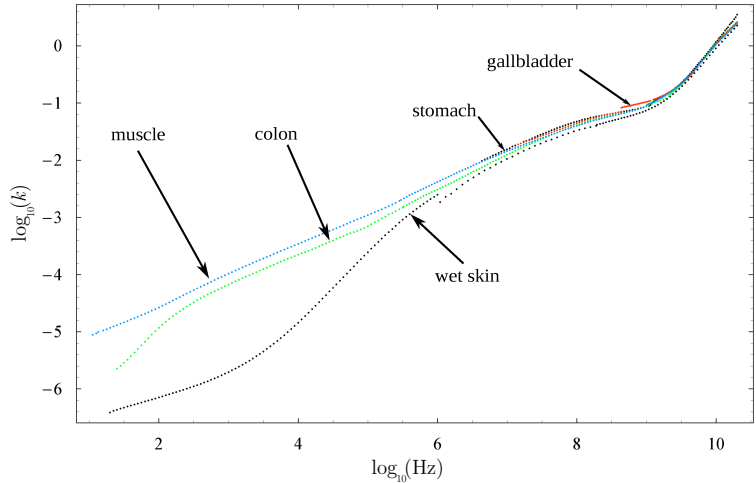}
\caption{Dielectric behavior of several tissue types after removing media conductivity.}\label{figure1}
\end{figure}

The approximation for the dielectric absorption for biological tissue, represented by equation (\ref{approx2}), was used to convert experimental conductivity measurements for several tissue types, so the same in the original analysis. The result of converting experimental conductivity measurements for several tissue types are displayed in figure (\ref{figure1}). After applying an appropriate quotient, the dielectric behavior of biological tissue is parsed out. The residual dielectric behavior aligns with theory and expectation, for the absorption curves shown in figure (\ref{figure1}) clearly show the effect of electrophoresis, the contraction of the thermal resonance peak, also, all tissue types conform to common expectations for dielectric materials.

Capitalizing on the transformation of experimental conductivity values for biological tissue, one may immediately notice that the approximate dielectric behavior of tissue mimics that of pure liquid water; yet, tissue does diverge from pure liquid water for very important reasons. The absorption exhibited by tissue is raised above that of pure liquid water, ranging from static fields to megahertz frequencies. Absorptive additions from proteins and the like are responsible for the increase in the extremely low frequency range; conversely, the increase within the megahertz range is primarily due to electrophoresis. Another reason absorption is raised and differentiated across tissue types is the presence of other dielectric materials, such as proteins, amino acids, \&c. The absorption rises monotonically to peak around 10 megahertz, which was predicted. Despite the range of ionic concentrations for different tissue types, the absorption behavior for a range of tissue types is seen to converge onto one another as the resonance frequency for electrophoresis is approached.

Another important feature clearly visible in figure (\ref{figure1}) is contraction of the absorption peak around the thermal maximum, where the thermal maximum represents energy loss due to rotational modes for bound water. Bound water is impeded in its free rotation by ionic electric fields; thus, shielding forces a reduction in absorption. The relaxation time for pure liquid water has been calculated to be roughly 5 picoseconds; in the case of tissue, it has been calculated to be around 17 picoseconds. The relaxation time for tissue has been dilated, indicating the presence of bound water. The relaxation time is approximated by measuring the Full Width Half Maximum (FWHM) for the absorption peak of interest. Because the relaxation time is dilated, a contraction of the thermal resonance peak is predicted by the Similarity Theorem 2.3, which is stated in the original analysis.

A variable penetration depth for tissue has been recorded in the optical frequency range \cite{Welch, LSM:LSM1134, Germer}. In the original analysis, an argument for measuring differing penetration depths was given, where the blame for the differences was attributed to experimental technique and numerical model employed to extract relevant dielectric properties. In retrospection, an obvious difference between the two main experimental techniques is the fact that one group measures tissue properties \emph{in vivo}, while the other group uses \emph{ex vivo} samples. The denaturation and coagulation of excised tissue is inevitable and leads immediately to the assertion that differing penetration depths are dependent upon tissue health, where Postulate \ref{coefficient_theory} embodies this assertion. As a consequence, the penetration depth is a function of external source parameters, such as power density; also, it is a function of material response to rising temperatures, which too is a function of time. Typical mappings of attenuation are seen on the order of $\mu_a:10^{-6}\mapsto 10^{-4}$. One contradiction still stands, Simpson \emph{et.\hspace*{-1pt} al.}\hspace*{-1pt} employed \emph{ex vivo} samples, yet the calibration method he devised appears to correct for this fact; see original analysis for details.

All three algorithms employed in the original analysis for interpolation, extrapolation and the combination thereto are given in Appendix \ref{code}. The combination of Neville's interpolation method with Richardson's extrapolation method is used to form a more accurate approximation from a set of discrete experimental data points. The only reason interpolation is entertained is that a regular set of data points is required for the entire interval transformed by the Kramers-Kr\"{o}nig relation. If multiple sets existed spanning all frequency bands, then arithmetic averaging would avail; but, for the most part, duplicate data sets do not exist nor is it acceptable to average data obtained from different experimental techniques, sample preparations and numerical methods designed to extract optical properties. 

The Kramers-Kr\"{o}nig transform is in actuality just Hilbert's transform and the most efficient means by which to emulate the Hilbert transform numerically over discrete sets is through Discrete Fourier transforms. The original analysis described in great detail the implementation of the Kramers-Kr\"{o}nig relation transforming a discrete absorption set to yield a theoretical index of refraction set, but if a more in-depth discussion is desired concerning the Hilbert transform, see the paper, ''Stieltjes Integral Theorem \& The Hilbert Transform'', published in 2011 \cite{2011arXiv1105.4179C}. 

\section{Discussion}
It was never the intent of either this essay nor the antecedent analysis to undermine reported conductivity for biological tissue, these values were determined by experiment; rather, the intent is to complement experimental measurements, by placing proper emphasis upon material behavior as supported by electromagnetic theory. Proper emphasis enables decoupling excess conductivity attributable to electrolytes and electric double layers from embedded dielectric materials present within biological tissue; moreover, the residual dielectric behavior revealed after decoupling conforms well with expectations based upon electromagnetic theory. 

Biological tissue is made primarily of liquid water, thus it seems natural to assume there to be a similarity between the two materials with respect to exhibited absorption characteristics. This assumption is based upon the well founded \emph{effective medium approximation}, of which, Maxwell's 'mixing rules' is a subordinate sub-theory. Heterogeneous mixtures are not expected to form properties completely unrelated to the materials that make it up; rather, it is elementary logic to expect the composite, effective or resultant properties to be a weighted sum of all the constituents present in the mixture. 

Heterogeneous materials pose a difficult problem when attempting to reconstruct and predict electromagnetic behavior from axiomatic principles. Maxwell's 'mixing rules' can often provide some means of adequate representation of the dielectric behavior for a mixture of materials; although, the mixing must be, in some sense, disjoint, as a result, one may consider the separate materials as being distinct but properly mixed \cite{Garnett}. A wider theory containing the concept of the 'mixing rules' is a theory called 'effective medium approximation', which states the \emph{effective} property of a heterogeneous compound is equal to the ratio of each material's intrinsic property comprising the compound \cite{Bruggeman}. 

This form of analysis has been applied in the optical frequency range, for at these frequencies, the excess conductivity caused by electric double layers is vanishingly small. Thus, it is common to see both experimenters and theoreticians attempt to deconstruct the resultant absorption properties of tissue in the optical frequency range. Such chemicals as hemoglobin, deoxyhemoglobin and melanin are all used to attempt a deconstruction of tissue's exhibited absorption characteristics. Attempting to apply an \emph{effective medium approximation} to the lower frequency range fails dismally, for in this frequency range, the dielectric properties of tissue are confounded by excess conductivity associated with electrolytes in solution and electric double layers surrounding biological cells. As can be seen, in the case of biological tissue, not only a mixing of materials do occur, but, in addition, the property of conduction pervades the entire media. 

Regardless of experimental results, biological tissue is classified as a dielectric material; therefore, excessively high conductivity measured for tissue proved initially nonplussed. Converting reported experimental conductivity values by the usual relation [\emph{sic}, equation (\ref{approximation})] produced results that were confusing, for the calculated absorption was found to increase towards the origin. It is by theory and classification that the absorption of a dielectric material should be vanishingly small for static fields. Equally troubling are the excessively high permittivity values indicating very large polarizability of tissue, which are some four to five orders of magnitude higher than pure liquid water. Mind you, water is a classic dielectric material, exhibiting very large polarization; it is difficult to reconcile the excessively large increase in polarizability for tissue simply because of the presence of electrolytes in solution. In fact, the presence of electrolytes actually depresses the polarizability in the case of saline solutions. Worse, if reported permittivity values are taken \emph{prima facie}, it is very difficult to attribute excessively large polarizability to the medium as a continuum. In other words, without attributing these large values to the cellular structure, specifically, the mesoscopic geometry, it becomes difficult to justify that the ionic media itself is causing such large polarizations. 

The original analysis performed found it necessary to transform radio frequency conductivity values for tissue in order to form a set of absorption values that conformed to electromagnetic theory and expectation. The cause for a transformation was placed squarely upon the shoulders of Maxwell-Wagner polarization phenomenon; furthermore, it was found necessary to add greater detail and theoretical support to uphold that original contention. This essay reaffirms the postulate that excess conductivity causes tissue to exhibit excessively high polarization and conductivity measurements; moreover, greater detail is given for why the excess conductivity hides the dielectric behavior of tissue.  

Conductivity in metals has been successfully described by ''electronic band theory'' and predicts a large band gap for insulators, hence, the reason for poor conduction in dielectrics \cite{Ashcroft}. Electrolytic solutions do not provide much geometric structure, hence, the resistivity would be predictably high from electronic band theory; yet, solvated electrons would lower the band gap, as evidenced by experiment, i.e. saline solutions. In the case of biological tissue, there is a regular geometric lattice, comprised of the cellular structure; therefore, the double layer surrounding each cellular membrane would lend yet another source decreasing the \emph{resultant} band gap. Ultimately, because of both electrolytic and double layer conduction, a conductive band permeates throughout biological tissue.

As far as the dielectric behavior of tissue is concerned, it would be expected to resemble pure liquid water based upon an \emph{effective medium approximation}. Modification of the resultant dielectric behavior should be predictable with the introduction of each new chemical species, such as ions, proteins, amino acids, \&c. Modification of the electromagnetic absorption behavior of biological tissue is shown to adhere to spectroscopic principles and expectations, including contraction of the thermal resonance peak, increased absorption in the optical spectrum and an additional absorption peak due to electrophoresis \cite{2010arXiv1010.3752C}. Many of these features are not visible before a quotient is applied to reported experimental conductivity values for biological tissue.

A spectroscopist would expect contraction of the absorption curve if told that the relaxation time for the material had dilated; this expectation is based upon Fourier theory, specifically, the Similarity Theorem 2.3 found in the original analysis \cite{2010arXiv1010.3752C}. In the case of biological tissue, the introduction of ions in solution would form what is referred to as 'bound water', that is, the electric potential associated with ions in solution will force water molecules to bind or become bound, see Corollary 2.4 found in the original analysis \cite{2010arXiv1010.3752C}. This means molecules are shielded from external fields by local electric fields emanating from ions in solution. Because of shielding from electrolytes in solution, the dielectric relaxation time for tissue is expected to dilate with respect to pure liquid water; consequently, contraction of the absorption curve is equally expected, i.e. the Similarity theorem.


\vspace*{10pt}\noindent \emph{\textbf{Copyright Statement}}\\
\emph{I am a military service member (or employee of the U.S. Government). This work was prepared as part of my official duties. Title 17 U.S.C. \S105 provides that 'Copyright protection under this title is not available for any work of the United States Government.' Title 17 U.S.C. \S101 defines a U.S. Government work as a work prepared by a military member or employee of the U.S. Government as part of that person's official duties.}


\bibliography{references}   
\bibliographystyle{spiejour}   

\newpage
\section{appendix: Mathematica codes}\label{code}
\noindent Three separate codes, written in Mathematica 5.2 from Wolfram Research, are listed below \cite{Mathematica}. 

\noindent 1. Combined interpolation/extrapolation algorithm:\\
\noindent\framebox[0.9\textwidth][c]{
\begin{minipage}{0.85\textwidth}{\scriptsize
InterpolationExtrapolation[xlistoo\_, resolution\_, rounds\_, richardsonorder\_] := \\
  Module[\{newdata, res = IntegerPart[resolution], up = rounds, i, p1, p2, order = richardsonorder, temp, x, xlisto\},\\
\hspace*{20pt} xlisto = Union[N[xlistoo], SameTest $\rightarrow$ (First[\#1] == First[\#2] \&)];\\
    newdata = List[];\\
    p1 = N[First[xlisto[[All, 1]]]];\\
    p2 = N[Last[xlisto[[All, 1]]]];\\
    baselist = Table[x, \{x, p1, p2, (p2 - p1)/(res - 1)\}];\\
    For[i = 1, i $<$= up, i++,\\
\hspace*{10pt}       newdata = Join[newdata, \{Table[ NevilleS[xlisto, x, (i + 1)],\\
\hspace*{20pt} \{x, p1, p2, (p2 - p1)/(res - 1)\}]\}]];\\
    temp = First[Richardson[Apply[List, Reverse[newdata]], order]];}
\end{minipage}}
\vspace{5pt}

\noindent 2. Neville's interpolation algorithm:\\
\noindent\framebox[.9\textwidth][c]{\scriptsize
\begin{minipage}{0.85\textwidth}{ NevilleS[xlisto\_, xint\_, len\_] := \\
Catch[Module[\{x = N[xint], q, data = N[xlisto], max, max2, i\},\\
Off[First::''first''];\\
Off[Last::''normal''];\\
max=n=2*len + 1;\\
q = Table[0, \{i, 1, n\}, \{j, 1, n + 1\}];\\
\{min2, max2\} =\{Min[Abs[(data[[\#1 + 1, 1]] - data[[\#1, 1]])]\&/@Range[1,Length[data]-1]],\\
\hspace*{10pt} Max[Abs[(data[[\#1 + 1, 1]] - data[[\#1, 1]])]\&/@Range[1,Length[data] - 1]]\}; i = 1;\\ 
\hspace*{20pt} While[x $>$= data[[i, 1]] \&\& i $<$ Length[data], num = i; i = i + 1];\\
\hspace*{30pt} If[IntervalMemberQ[Interval[\{Min[data[[All, 1]]], Max[data[[All, 1]]]\}], x] \&\& NumberQ[num],\\
\hspace*{10pt} If[IntervalMemberQ[Interval[\{1, len\}], num] $\vert\vert$\\
\hspace*{20pt} IntervalMemberQ[Interval[\{Length[data] - len, Length[data]\}], num],\\
\hspace*{10pt} If[IntervalMemberQ[Interval[\{1, len\}], num], q[[All, 1]] = Take[data[[All, 1]], n];\\
\hspace*{20pt} q[[All, 2]] = Take[data[[All, 2]], n], q[[All, 1]] = Take[data[[All, 1]], -n];\\
\hspace*{20pt} q[[All, 2]] = Take[data[[All, 2]], -n]], q[[All, 1]] =\\
\hspace*{30pt} Take[data[[All, 1]], \{IntegerPart[num] - len, IntegerPart[num] + len\}];\\
q[[All, 2]] = Take[data[[All, 2]], {IntegerPart[num] - len, IntegerPart[num] + len}]],\\
\hspace*{30pt} Throw[''Error: out of bounds'']];\\
For[i = 2, i $<$= n, For[j = 1, j $<$= i - 1,\\
\hspace*{20pt} q[[i, j+2]] = ((x - q[[i - j, 1]])*q[[i, j + 1]] -\\
\hspace*{30pt} (x - q[[i, 1]])*q[[i - 1, j + 1]])/(q[[i, 1]]-q[[i-j,1]]);\\
\hspace*{40pt} j++]; i++]; Last[Last[q]]]];}
\end{minipage}}
\vspace{5pt}

\noindent 3. Richardson Extrapolation algorithm:\\
\noindent\framebox[0.9\textwidth][c]{
\begin{minipage}{0.85\textwidth}{\scriptsize
 Richardson[x\_\_, order\_] := \\
 \hspace*{10pt} Module[\{p, temp = x\}, For[i = 1, Length[x] $>$ i, i++, \\
        \hspace*{20pt} temp = Map[((order\textasciicircum(i))*temp[[\#1 + 1]] - temp[[\#1]])/((order\textasciicircum(i)) - \\
                    \hspace*{30pt} 1) \&, Range[1, Length[temp] - 1]]]; temp];}
\end{minipage}}
\end{document}